# Valence band and core-level analysis of highly luminescent ZnO nanocrystals for designing ultrafast optical sensors

Amish G. Joshi, Sonal Sahai, Namita Gandhi, Y. G. Radha Krishna and D. Haranath[a]

*National Physical Laboratory, Council of Scientific and Industrial Research, Dr. K. S. Krishnan Road, New Delhi – 110 012, India.*

Highly luminescent ZnO:Na nanocrystals of size ~2 nm were synthesized using a improved sol-lyophilization process. The surface analysis such as survey scan, core-level and valence band spectra of ZnO:Na nanocrystals were studied using x-ray photoelectron spectroscopy (XPS) to establish the presence of $Na^+$ ions. The observed increase in band gap from 3.30 (bulk) to 4.16 eV (nano), is attributed to the quantum confinement of the motion of electron and holes in all three directions. The photoluminescence and decay measurements have complemented and supported our study to design an efficient and ultrafast responsive optical sensing device.

[a]Electronic mail: haranath@nplindia.org

Sensors or detectors for various kinds of optical radiations are known in the literature[1,2], but they are deficient in many ways. For example, the photo-response of known semiconductor sensors in ultraviolet (UV) and x-ray radiation range is limited due to non-radiative surface recombinations. The radiation with energy greater than few hundred meV above the band gap will be absorbed ~10 nm deep near the surface. The photo-generated free carriers due to the absorption recombine non-radiatively near the surface of the semiconductor[3]. However, in the visible/infrared region the absorption length of the radiation is significantly larger and the recombination occurs deeper in the material away from the surface. The carrier recombination at the surface could be avoided either chemically passivating the surface or using heterojunction schemes, which is evidenced in many III-V semiconductors devices[3]. But the above two schemes fail if the incident radiation is of very high energy such as UV, x-ray, synchrotron radiation etc. Hence, there is a serious pursuit globally for newer materials that are radiation sensitive, have higher brightness that should match commonly-used photodetector, minimum self-absorption, reduced light scattering and very short recombination time, suitable for designing ultrafast optical sensors[4]. Phosphors are the best materials to detect UV radiation, but their practical usage for sensing has been limited due to their longer decay/lifetimes and light scattering. However, nanophosphors of the same kind could very well satisfy the prerequisites of an optical sensor due to their highest brightness levels and shortest possible lifetimes dictated by their sizes. A combination of an efficient nanophosphor and a known optoelectronic detector could yield an advanced optical sensor with increased signal to noise ratio, is the main objective of our work. High brightness ZnO nanophosphors having size ~2 nm were the best choice in the current study as they exhibit two luminescence bands in the UV and green regions of the electromagnetic spectrum[5]. This makes the material versatile to be used to design an efficient optical radiation detector covering a broad wavelength range. UV emission from ZnO is caused by band-to-band and exciton transitions[6] whereas; the origin of green photoluminescence (PL) band is still on debate[7-10]. Spatial confinement of charge carriers could be the possible reason for substantial increase in the photoluminescence.[11] Considering this property in view we thought it is interesting to investigate the surface states of alkali doped-ZnO nanocrystals (NCs) by surface sensitive spectroscopic techniques and contribute to design an ultrafast optical sensing device that is operable in the nanosecond time scale.

In a typical experiment, an improved sol-lyophilization process[12] was used for the synthesis of $Na^+$ doped ZnO NCs. This method allows for the synthesis of ultra-fine nanocrystals with no surfactant. The details and the associated advantages of the process are described elsewhere.[13] In a typical method of preparation, ethanolic stock solutions of the reactants, zinc acetate and sodium hydroxide (NaOH) were allowed to react at room temperature (~25°C). The nucleation and growth rate of the ZnO NCs were monitored by adjusting the pH, relative humidity and the pervading temperatures. Immediately after the reaction, the nanocrystal agglomerates were precipitated out by adding a low relative polarity solvent such as n-hexane, washed and dried below 50°C. In another experiment, calculated amount of high (99.9%) purity ZnO powder procured from Sigma-Aldrich was mixed well with NaOH and annealed in air for 5 hours at 1200°C in order to form a standard bulk-ZnO:Na sample. These two representative bulk and nano-ZnO:Na samples were chosen for the present study.

Figure 1 shows the x-ray diffraction (XRD) pattern of the ZnO:Na NC powder, which shows that the particles are ultra-fine and have wurtzite phase. The inset of figure 1 portray the transmission electron microscopy (TEM) image taken at a magnification of 150 kX and the photograph of a vial containing highly luminescent ZnO:Na NC powder kept under UV (350 nm) radiation. From the broadening of XRD lines and TEM image, the average sizes of NCs were estimated to be ~1.5 and ~1.7 nm, respectively, after duly subtracting the XRD profile broadening due to lattice strain.

Incorporation of $Na^+$ in ZnO samples were initially confirmed by atomic absorption spectroscopy, which is a powerful tool to determine quantitative elemental composition of $Na^+$ for bulk and nano-ZnO:Na phosphors and hence, were estimated to be 12.31 and 10.25%, respectively, with an error of ±2%. Further, the values were compared to the results from inductively coupled plasma atomic emission spectroscopy reported by Kshirsagar *et al*.[14]. Using x-ray photoelectron spectroscopy (XPS) and core-level spectra we tried to characterize the surface of ZnO:Na samples thoroughly at ~5 × $10^{-8}$ torr with a non-monochromatized Al Kα x-ray source. The pass energies for survey scan and core level spectra were kept at 100 and 60 eV, respectively. The minute specimen charging observed during photoemission studies was calibrated by assigning the C 1s signal at 285 eV taking an internal reference and the Fermi edge of gold sample. Survey spectra in figure 2 shows sharp peaks of C 1s (284.6 eV), O 1s (531 eV), Zn ($2p_{3/2}$) and Zn ($2p_{1/2}$) at 1021 and 1043 eV; whereas peaks related to Na were observed at Na(1s) (1071 eV), Na(2s) (64 eV) and Na(2p) (31 eV). In addition, weak peaks were observed for NCs compared to bulk samples that indicate poor crystallinity of the former. The detailed study on the core level XPS spectra of Zn, O and Na, shown in figure 3(a)-(c), have shown symmetric profiles depicting uniform bond structure and emphasizes that both bulk and NC's do exist in same phase. The core level spectra (Fig 3(a)) of Zn shows two distinct Zn($2p_{3/2}$) and Zn($2p_{1/2}$) states at 1020.9 (nano) and 1021.65 eV (bulk); and at 1043.9 (nano) and 1044.4 eV (bulk) respectively. The separation of the states in NCs and bulk is 23 and 22.75 eV respectively, which confirms the phase formation of ZnO aroused due to the spin-orbit splitting of energy levels. There could be slight shifts in binding energies due to variation from site to site substitution of Na in ZnO systems. Figure 3(b) shows O(1s) core level spectra. The peaks X and X′ observed at 531.75 and 532.15 eV are the contributions from oxygen of ZnO and partially from Na-O-Na bonding. Interestingly there is an appearance of shoulder at higher binding energies (536 eV), which is denoted as Y and Y′ for bulk and nano-phases, which is attributed to the presence of OH group at ZnO surface.[15] The peak intensity of shoulder peak Y′ is higher than that of Y, which clearly manifests the adsorption of more OH groups in nanophase due to larger surface to volume ratio[16]. Our results agree very well with *Ab initio* calculations[15] where X and X′ are related to bridging oxygen, whereas Y and Y′ to central oxygen of ZnO structure. Figure 3(c) shows Na (1s) high resolution spectra of bulk and nano-phases which appear almost symmetric and further establishes the presence of Na$^+$ ions.

Figure 4 shows the valence band (VB) XPS spectra of ZnO:Na for bulk and NCs acquired at a pass energy of 60 eV. The weak peaks around Fermi level, $E_F$, are estimated to be a kind of non-local state in the band gap possibly caused by defects. Peak A and A′ observed at ~6 eV in bulk and NCs mainly involves O $2p$ orbitals and the part of O $2p$ orbitals hybridized with Zn $4s$ and $4p$ ones. Peak B and B′ for bulk and nano observed at 11 and ~10.5 eV respectively, is mainly attributed to the Zn $3d$ band. Our valance band spectra analysis matches very well with the earlier reports of ZnO systems.[17,18] In NC, the Zn $3d$ moves to lower energy by ~0.5 eV with respect to bulk. In peak regions of A and A′, the uppermost valence-band edges (VBE) could be easily separated out. The VBE was linearly fitted to determine the shift, $\Delta E_v$, as shown by diagonal lines in figure 4. The tangent to the fitted curve was drawn and intersect is defined as VBM. The binding energy shift in VBM with respect to $E_F$ is obtained. It is evident from figure 4 that the VBM of nano-ZnO shifts towards higher binding energy by ~0.42 eV than bulk-ZnO. The band gaps estimated are 3.30 and 4.16 eV for bulk and nano-ZnO:Na, respectively, which are in excellent agreement with the values derived from XRD and TEM observations.

Since it is well known that spectroscopic methods such as optical absorption, photoluminescence excitations etc. are accurate to determine the crystallite band gap, we performed detailed luminescence studies. Figure 5 shows the typical photoluminescence excitation (PLE) and photoluminescence (PL) spectra of NCs and bulk phosphor samples. PL studies showed a clear superiority of brightness levels in ZnO:Na NCs over their bulk counterpart. The ZnO:Na NCs synthesized by the present method exhibit a distinct excitation peak in the UV region at ~302 nm (4.15 eV), when monitored at 539 nm, which is a clear indication of smaller size of the particles. However, the bulk phosphor having larger crystallite sizes, when monitored at 545 nm, exhibited the excitation peak around 375 nm (3.3 eV) that corresponds to bulk band gap of ZnO system. The difference in PLE peak position (or the band edge) for NC and bulk-ZnO:Na samples is mainly due to different mean sizes. The broad green emission at ~545 nm is weakly observed in the bulk-ZnO:Na and is assigned to defect levels associated with oxygen vacancies or zinc interstitials,[7,8] whereas the confinement effects dominate the luminescence phenomenon in ZnO NCs. There could be many interdependent factors, such as electron-phonon coupling, lattice dislocation, localization of charge carriers due to interface effects and point defects that are incrementing the PL emission. However, the sizes observed for ZnO NCs are ~2 nm, the NCs allow one more degree of freedom along with size dependency leading to intense PLE and PL intensities over their bulk counterparts. This is the property related to the sensitivity where most of the optical sensors lack in the UV to visible region. It is known that the performance of an optical sensor would be optimum if the incident photon energy is transferred at a faster rate from host crystal to dopant. To check the possibility of using ZnO:Na NCs for designing ultrafast UV sensor, we have investigated the dynamics of bound excitons for bulk and nano-ZnO:Na powders at their peak wavelengths i.e., at 545 nm emission and 375 nm excitation for bulk; and 539 nm emission and 302 nm excitation for NCs, using a time correlated single photon counting technique. Table I shows the results of exciton lifetimes of both bulk and nano-ZnO:Na powders that varied from few milli to nanoseconds, respectively. The lifetime components (fast and slow) are calculated at the $1/10^{th}$ part of the initial decay curves for bulk and ZnO:Na NCs and the values are shown in Table I. It is inferred that size dependency and quantum confinement are responsible for faster decay components and donor-acceptor pair type of recombination for longer lifetimes, are for bulk materials.

In summary, highly luminescent ZnO:Na nanocrystals were investigated for designing ultrafast optical sensor for which the surface states of NCs were studied using XPS. The valence band maximum of nano-ZnO:Na shifts by ~420 meV from Fermi energy level as compared to that of bulk-ZnO:Na, which is due to the quantum confinement effects aroused due to size restrictions in nanocrystals. The fastest rate of energy transfer from host to the activator in ZnO:Na NCs suggest the strong possibility of designing advanced UV sensor that could be operated in nanoseconds.

The authors (NG, YGRK) acknowledge CSIR and UGC, India for their fellowships.

**Table I**

The results pertaining to exciton lifetimes of both bulk and nano-ZnO:Na powders are shown below. The lifetime components (fast and slow) are calculated at the $1/10^{th}$ decay of the initial brightness for all the samples.

| S. No. | Size determination using | | | PLE peak wavelength | Exciton lifetime components | | Reference |
|---|---|---|---|---|---|---|---|
| | XRD | TEM | PLE | | Fast | Slow | |
| 1 | 1.5 nm | 1.7 nm | 1.9 nm | 302 nm | 5.6 ns | 17.5 ns | This study |
| 2 | 2.5 nm | 2.2 nm | 2.9 nm | 332 nm | 35.4 ns | 569 ns | [10] |
| 3 | 3.5 nm | 3.9 nm | 3.7 nm | 342 nm | 182 ns | 1255 ns | [10] |
| 4 | 7.6 nm | 7.9 nm | 8.0 nm | 351 nm | 1660 µs | 424 µs | [10] |
| 5 | >20 µm | >20 µm | >20 µm | 375 nm | 325 ms | 1880 ms | This study |

**Figure captions**

FIG.1. Powder XRD pattern of ZnO:Na NC's. Insets show the TEM micrograph and a vial containing ZnO:Na NC powder irradiated under UV (350 nm) light.

FIG. 2. XPS survey scan spectra of bulk and NC's of ZnO:Na recorded with a photon energy of Al $K\alpha$ ($h\nu$=1486.6 eV).

FIG. 3. Core level XPS spectra of (a) Zn (b) O and (c) Na in bulk and NC's of ZnO:Na.

FIG. 4. VB XPS spectra of bulk and NC's of ZnO:Na acquired at the pass energy of 60 eV.

FIG.5. Room temperature photoluminescence (PL) and photoluminescence excitation (PLE) spectra for bulk and NC's of ZnO:Na.

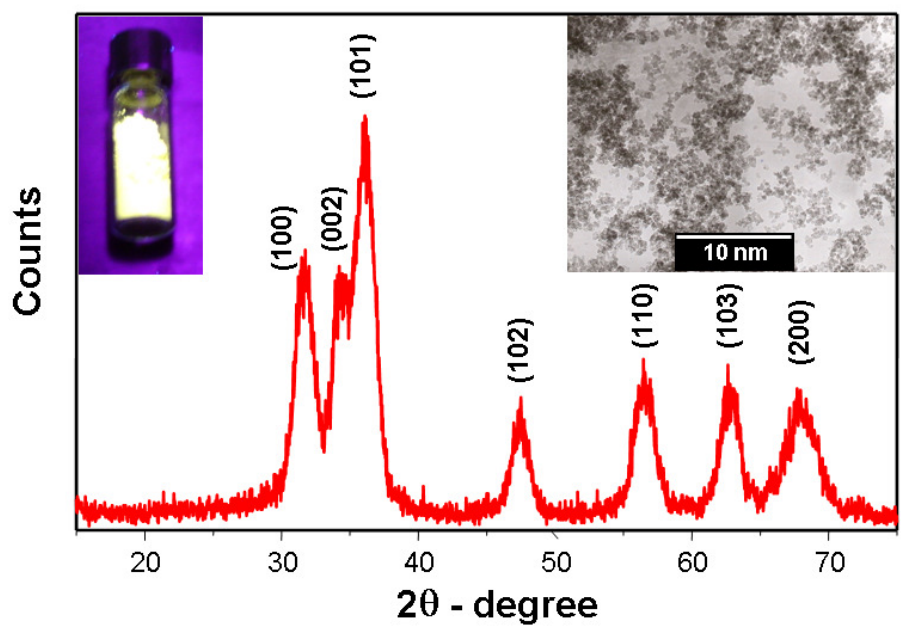

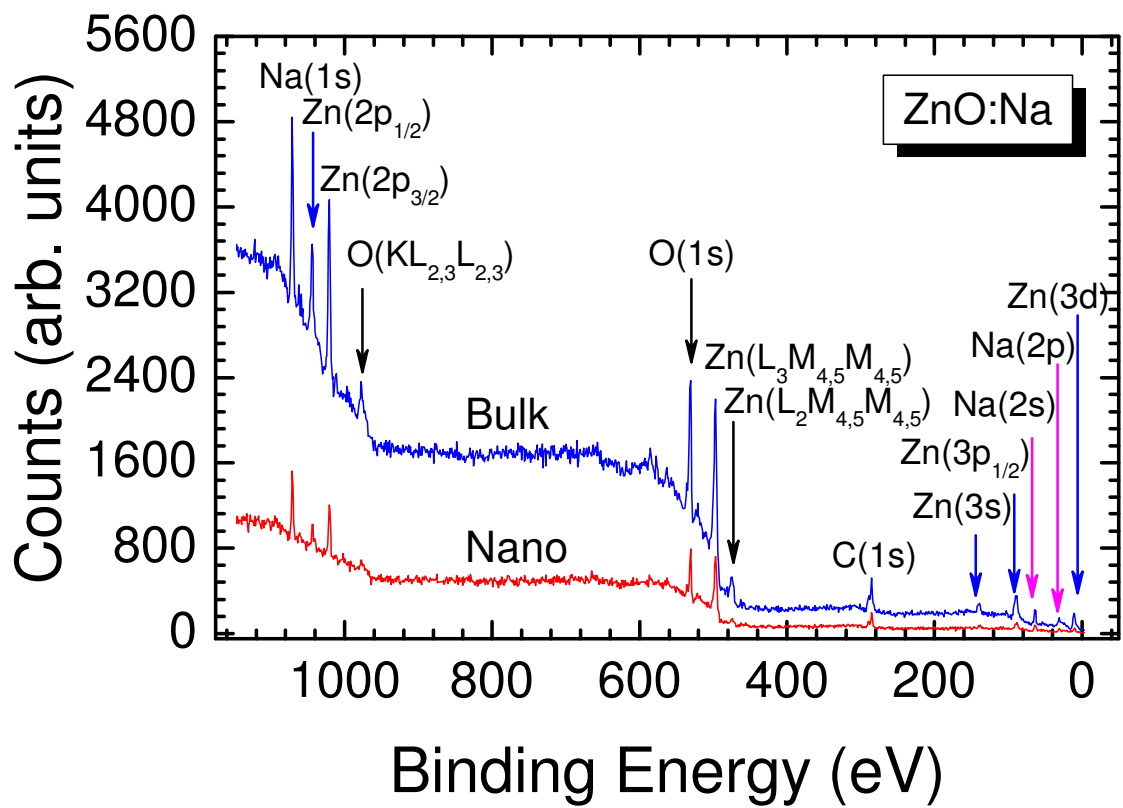

Fig.2

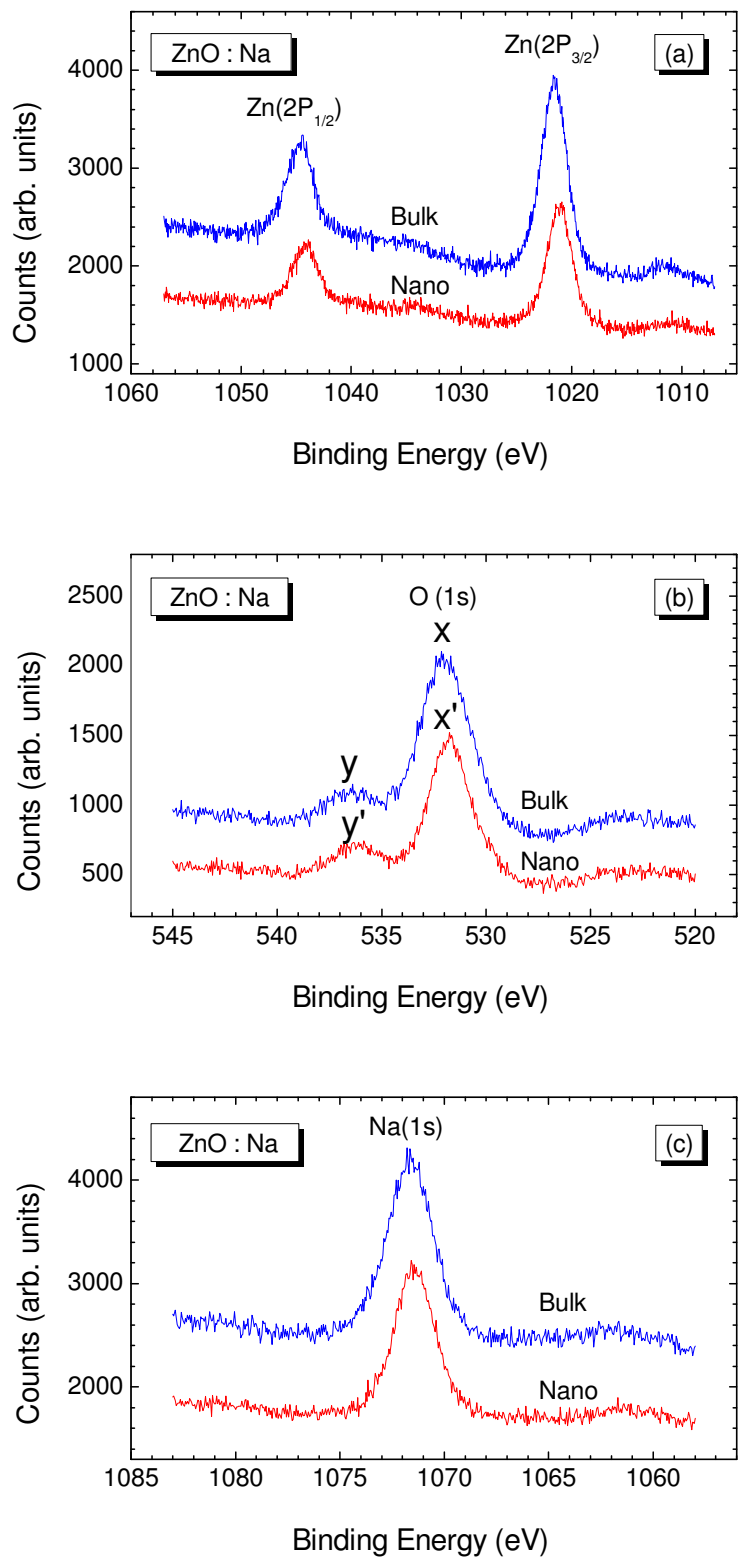

Fig. 3

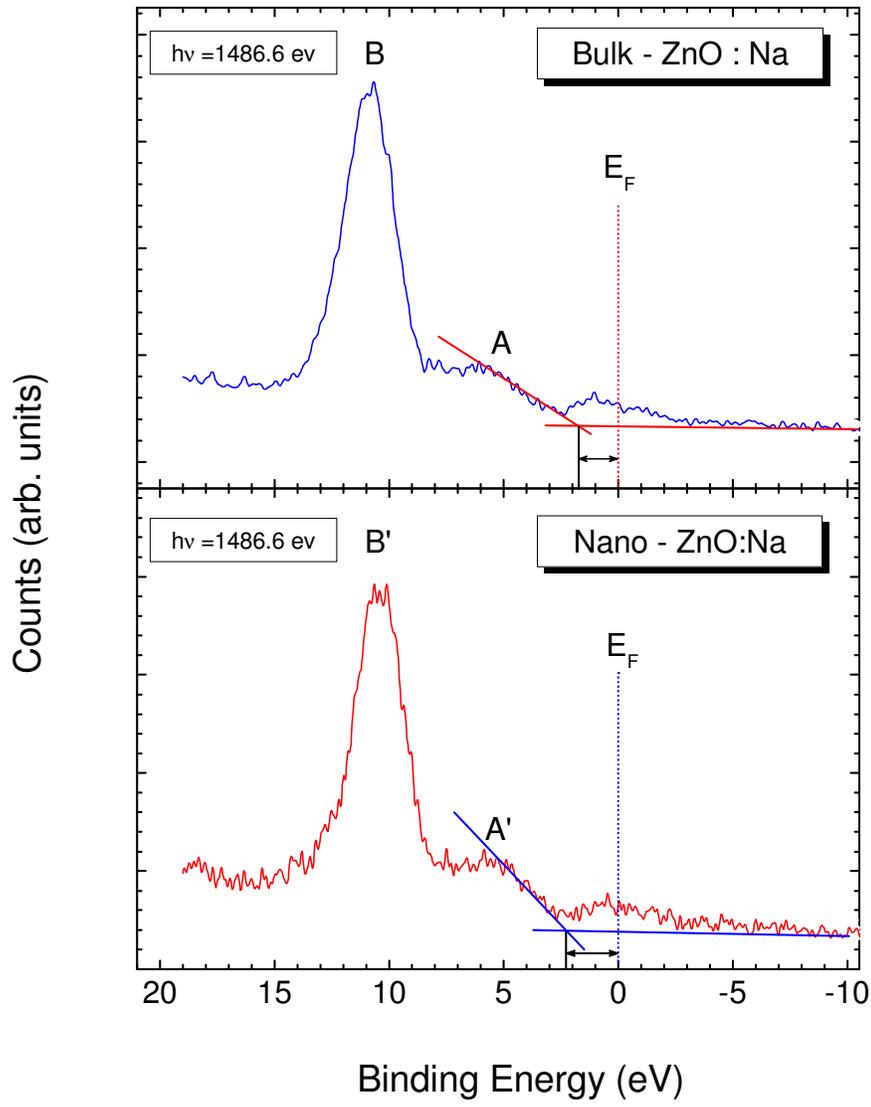

Fig. 4

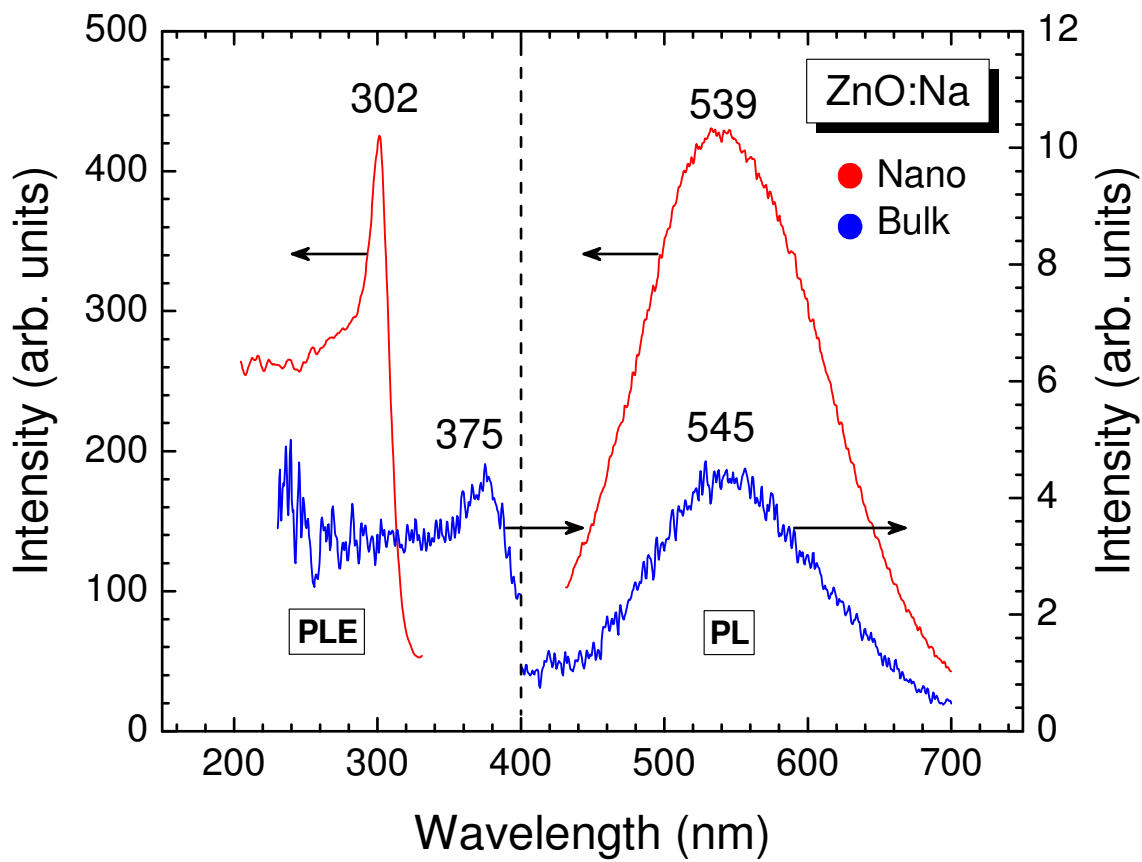